\documentclass[english,12pt]{article}
\usepackage{array}
\usepackage{graphicx}
\usepackage{amssymb}
\usepackage{amsmath}
\usepackage{multirow}
\usepackage{prettyref}
\usepackage{babel}
\usepackage{units}
\usepackage[latin1]{inputenc}
\usepackage{amsfonts}
\usepackage{amssymb}
\usepackage{babel}
\usepackage{subcaption}
\usepackage{cite}
\usepackage{slashed}
\usepackage{booktabs}
\usepackage[colorlinks,pdfusetitle,urlcolor=blue,citecolor=blue,linkcolor=blue,bookmarksnumbered,plainpages=false]{hyperref}

\def\@fmsl@sh#1#2#3{\m@th\ooalign{$\hfil#1\mkern#2/\hfil$\crcr$#1#3$}}
 \def\eq#1\en{\begin{equation}#1\end{equation}}
\def\s[#1,#2]{[#1\stackrel{\star}{,}#2]}
\def\sx[#1,#2]{[#1\stackrel{\star_{x}}{,}#2]}

\DeclareMathOperator{\Tr}{Tr}

\newcommand{\nc}{\newcommand}
\nc{\beq}{\begin{equation}}
\nc{\eeq}{\end{equation}}
\nc{\beqa}{\begin{eqnarray}}
\nc{\eeqa}{\end{eqnarray}}

\def\bc{\begin{center}}
\def\ec{\end{center}}

\def\to{\rightarrow}

\def\gsim{\mathrel{\mathpalette\atversim>}}

\def\bc{\begin{center}}
\def\ec{\end{center}}

\def\gsim{\mathrel{\rlap{\lower4pt\hbox{\hskip1pt$\sim$}}

    \raise1pt\hbox{$>$}}}       

\def\gsim{\mathrel{\rlap{\lower4pt\hbox{\hskip1pt$\sim$}}
    \raise1pt\hbox{$>$}}}       

\begin{document}
\makeatletter
\def\fmslash{\@ifnextchar[{\fmsl@sh}{\fmsl@sh[0mu]}}
\def\fmsl@sh[#1]#2{%
  \mathchoice
    {\@fmsl@sh\displaystyle{#1}{#2}}%
    {\@fmsl@sh\textstyle{#1}{#2}}%
    {\@fmsl@sh\scriptstyle{#1}{#2}}%
    {\@fmsl@sh\scriptscriptstyle{#1}{#2}}}
\def\@fmsl@sh#1#2#3{\m@th\ooalign{$\hfil#1\mkern#2/\hfil$\crcr$#1#3$}}
\makeatother

\thispagestyle{empty}
\begin{titlepage}
\boldmath
\begin{center}
  \Large {\bf One-loop corrections to $\eta/s$ in AdS$_4$/CFT$_3$}
    \end{center}
\unboldmath
\vspace{0.2cm}
\begin{center}
{{\large Iber\^e Kuntz$^{abc}$}\footnote{kuntz@bo.infn.it}, {\large Rold\~ao da Rocha$^a$}\footnote{roldao.rocha@ufabc.edu.br}}
 \end{center}
\begin{center}
{\sl $^a$
Federal University of ABC, Center of Mathematics,
\\
09210-580, Santo Andr\'e, Brazil}
\\
$ $
\\
{\sl
$^b$
Dipartimento di Fisica e Astronomia, Universit\`a di Bologna,
\\
via Irnerio~46, I-40126 Bologna, Italy}
\\
$ $
\\
{\sl 
$^c$
I.N.F.N., Sezione di Bologna, IS - FLAG
\\
via B.~Pichat~6/2, I-40127 Bologna, Italy
}
\end{center}
\vspace{5cm}
\begin{abstract}
\noindent We study quantum corrections at one-loop order to the shear viscosity to entropy ratio by implementing the Vilkovisky-Barvinsky effective action in asymptotically Anti-de Sitter spacetimes. The shear viscosity is shown to receive no corrections at this order, but the entropy acquires a logarithmic correction. The coefficient of this logarithm turns out to depend on the spin of the particles running in the loop and it can be either positive or negative. On the basis of this result, we argue that the Kovtun--Son--Starinets bound cannot be seen as a fundamental property of nature beyond the classical regime.
\end{abstract}  
\end{titlepage}



\newpage

\section{Introduction}
Dualities in physics have become a major subject of investigation since the prominent work by Maldacena \cite{Maldacena:1997re}. His foundational paper sets the ground for the equivalence between gauge theory and gravity, which led to a breakthrough in theoretical physics, as it promoted a novel way of relating different fields that turned out to be not so different at all. This intriguing conjecture, although having its roots in string theory, has become one of the most popular ideas in high-energy physics (see \cite{Natsuume:2014sfa,Hubeny:2014bla} for an overview on the topic).

Although theoretically very appealing, the AdS/CFT correspondence has not made its way into the real world yet possibly due to various top-down features, such as supersymmetry, that have not been realized in the laboratory. However, the observation of the shear viscosity to entropy ratio $\eta/s$ of the quark-gluon plasma measured in the Relativistic Heavy Ion Collider (RHIC) is close to the predicted value, thus providing support for a strongly coupled plasma above the QCD phase transition. The difference of the observed value with respect to the prediction is attributed to the fact that QCD does not possess the all the required properties that would make AdS/CFT work. Furthermore, the correspondence is established for $N\to\infty$, which conflicts with the finite number of colors $N=3$ in QCD. Deformations on the gauge side of the correspondence allows one to make progress towards more realistic scenarios, including the finite $N$ case. This corresponds to quantum corrections on the gravitational side.

In this paper, we look at one-loop corrections to general relativity in AdS$_4$ and study their effects on the corresponding three-dimensional gauge theory. We take an effective field theory approach to investigate the model-independent infrared portion of quantum general relativity \cite{Donoghue:1994dn,Burgess:2003jk}. This differs from the usual local higher curvature corrections which requires knowledge of the UV and thus corresponds to the unreliable, from a bottom-up perspective, high-energy part of loops\footnote{Note that the non-renormalizability of general relativity must be dealt with in the spirit of effective field theory, which is in principle valid up to the Planck scale $M_p$. In dimensional regularization, momentum integrals of loop diagrams extends to infinity, allowing for the calculation of UV divergences. These divergences come from the high-energy domain ($\gsim M_p$) of loop integrals, where the effective theory breaks down. Therefore, the form of the divergences (and consequently everything that is calculated from them) are Planckian effects and cannot be predictions of the low-energy theory. That is why they are unreliable in a bottom-up construction. Their only role at low energies is to parametrize our ignorance of the UV in terms of free parameters.}. It is important to make a distinction between our bottom-up construction, typical of effective theories, and the top-down quantum corrections due to the genus expansion commonly found in the literature \cite{Buchel:2004di,Buchel:2008ac,Buchel:2008wy,Myers:2008yi,Buchel:2008ae,Garousi:2008ai,Ghodsi:2009hg,Buchel:2008sh,Cohen:2007qr,Son:2007xw,Brigante:2007nu,Brigante:2008gz,Kats:2007mq,casadio2016,Ferreira-Martins:2019wym}. In the former, local terms encode in their coefficients the unknown high-energy information, which must be fixed by observations at some energy scale rather than computed from first principles as in the latter. Our primary interest is, however, in the infrared portion of quantum gravity, where both quantum general relativity and string theory should agree.

We are mainly interested in the shear viscosity to entropy ratio $\eta/s$ as it is one of the most important observables used to corroborate the AdS/CFT correspondence. We show that the shear viscosity does not receive corrections at one-loop order. The entropy, on the other hand, is modified by a logarithm term. This logarithmic correction has already been identified in other contexts \cite{Banerjee:2008fz,Cai:2009ua,Aros:2010jb,Banerjee:2010qc,Banerjee:2011jp,Sen:2012cj,Sen:2012dw,Kaul:2000kf,Carlip:2000nv,Govindarajan:2001ee,Mukherji:2002de,Solodukhin:1994st,Fursaev:1994te,Mann:1997hm,El-Menoufi:2015cqw}. Due to the spin dependence of the quantum action, we argue that the ratio $\eta/s$ violates the Kovtun--Son--Starinets (KSS) bound even within quantum general relativity. Violations of the KSS have also been observed in modified gravity \cite{Buchel:2004di,Buchel:2008ac,Buchel:2008wy,Myers:2008yi,Buchel:2008ae,Garousi:2008ai,Ghodsi:2009hg,Buchel:2008sh,Cohen:2007qr,Son:2007xw,Brigante:2007nu,Brigante:2008gz,Kats:2007mq}, including Horndeski gravity \cite{Feng:2015oea}. Such violations, however, were obtained at the classical level and result from additional degrees of freedom and/or interactions. As one approaches scales close to $\hbar$ in the gravitational theory, loop corrections become important and an accurate assessment of the KSS bound requires the inclusion of such quantum corrections. In fact, recall that quantum gravity corrections correspond to finite $N$ corrections on the gauge side of the correspondence, thus loop corrections in the gravity side are unavoidable to study real systems such as the quark-gluon plasma. This naturally raises the question of whether these violations to the KSS bound at the classical regime would survive the corrections imposed by quantum gravity. We stress that, in the present paper, no modification of general relativity is performed. Our focus is rather on the quantum corrections to general relativity, systematically obtained in the effective field theory of gravity.

This paper is organized as follows. In Sect.~\ref{sec:action}, we review the general formalism due to Vilkovsky \textit{et al} for the construction of a gauge-independent quantum action for gauge theory and quantum gravity. Sect.~\ref{sec:eta} is devoted to the calculation of one-loop corrections to the shear viscosity of the gauge theory. In Sect.~\ref{sec:s}, we employ the Euclidean formalism to calculate quantum corrections to the entropy of the Schwarzschild-AdS (SAdS) black hole, finally leading to the ratio $\eta/s$. We then assess and discuss our results in Sec.~\ref{sec:conc}.

\section{Vilkovisky-Barvinsky effective action}
\label{sec:action}
The quantum action is a central object in quantum field theory. It contains, in a single place, all the information regarding correlation functions, which is of utmost importance for experimental physics, and yet provides the dynamics of the mean field due to the backreaction of quantum modes. A general scheme for the computation of the quantum action in arbitrary backgrounds for a gauge theory of arbitrary spin has been introduced in \cite{Barvinsky:1985an,Barvinsky:1987uw,Barvinsky:1990up,Barvinsky:1993en}. In the following, we give a brief overview of this formalism.

In the background field formalism, the quantum effective action is determined by the following functional integro-differential equation in the Euclidean formalism
\begin{equation}
e^{-\Gamma[\phi]} = \int\mathcal{D}\Phi \exp^{-S[\Phi] + \int\mathrm{d}x (\Phi(x)-\phi(x)) \frac{\delta\Gamma[\phi]}{\delta\phi}},
\label{eq:pathint}
\end{equation}
where $S[\Phi]$ comprises the classical Einstein-Hilbert action in the presence of arbitrary matter fields $S_m$ and counter-terms $S_{ct}$ used for renormalization
\begin{equation}
S[\Phi] = -\int\mathrm{d}^4x\sqrt{-g} \frac{1}{16\pi G}(R-2\Lambda) + S_m + S_{ct},
\end{equation}
where $G$ is the Newton's constant, which will be set to unity in the next sections. At one-loop level, the divergences are proportional to terms containing up to fourth-order derivatives \cite{tHooft:1974toh}, thus the counter-terms will be given by
\begin{equation}
S_{ct} = \int\mathrm{d}^4x\sqrt{-g}\left[a_1 R^2 + a_2 R_{\mu\nu}R^{\mu\nu} + a_3 R_{\mu\nu\rho\sigma}R^{\mu\nu\rho\sigma} + a_4 \Box R\right],
\end{equation}
where the $a_i$ are bare coefficients. The path integral is over the quantum field $\Phi$ and depends parametrically on the mean field $\phi(x) = \left<\Phi(x)\right>$. The notation $\Phi=\Phi^A(x)$ is used to denote collectively fields of arbitrary spin. To perform the above path integral, one writes $\Gamma$ as a loop expansion
\begin{equation}
\Gamma[\phi] = \sum_n \Gamma^{(n)}[\phi],
\end{equation}
where $\Gamma^{(0)}$ denotes the classical Einstein-Hilbert action with a cosmological constant and $\Gamma^{(1)}$ is the one-loop contribution, which is given by
\begin{align}
&\Gamma^{(1)} = \frac12 \log\det F(\nabla) = \frac12 \Tr\log F(\nabla),\\
&F(\nabla)\delta(x,y) = \frac{\delta^2 S[\phi]}{\delta\phi(x)\delta\phi(y)}.
\end{align}
Using the Schwinger proper time method, the one-loop contribution to the quantum action can be written as
\begin{equation}
\Gamma^{(1)} = -\frac12\int_0^\infty \frac{\mathrm{d}s}{s}\Tr K(s),
\end{equation}
where $K(s) = e^{sF(\nabla)}$ is the heat kernel. The calculation of $\Gamma^{(1)}$ relies on approximate solutions to the heat equation satisfied by $K(s)$. The Schwinger--DeWitt method, for example, consists of a time asymptotic expansion of the heat kernel $K(s)$ at small $s$. Our interest is the covariant perturbation theory approach, whose approximation scheme is an expansion in powers of the curvature.

The operator $F(\nabla)$ can generically be written as
\begin{equation}
F(\nabla) = \Box + \hat P - \frac16 R,
\end{equation}
where $\Box = g^{\mu\nu}\nabla_\mu\nabla_\nu$ and $\hat P$ is an arbitrary potential term. The metric $g_{\mu\nu}$ and the connection $\nabla_\mu$ are characterized by the Riemann curvature $R^{\mu\nu\rho\sigma}$  and the fiber bundle curvature $\mathcal R_{\mu\nu} = \mathcal R^A_{\ B\mu\nu}$, respectively:
\begin{align}
[\nabla_\mu,\nabla_\nu]V^\alpha = \mathcal{R}^\alpha_{\ \beta\mu\nu}V^\beta,\\
[\nabla_\mu,\nabla_\nu]\Phi^A = \mathcal{R}^A_{\ B\mu\nu}\Phi^B,
\end{align}
for some vector field $V^\alpha$. We denote by $\Re = \{\hat P,R^{\mu\nu\rho\sigma},\mathcal R_{\mu\nu}\}$ the set of curvatures that characterizes the operator $F(\nabla)$. The purpose of covariant perturbation theory is to obtain all quantities of interest as an expansion in $\Re$. After a lengthy calculation, the resultant quantum effective action in arbitrary dimensions $2\omega$ to second order in curvature reads
\begin{align}
    \Gamma[\phi]=-\frac{\varGamma(2-\omega)\varGamma(\omega+1)\varGamma(\omega-1)}
    {2(4\pi)^\omega\varGamma(2\omega+2)}\int dxg^{1/2}(x){\rm tr}\Big\{R_{\mu\nu}(-\Box)^{\omega-2}R^{\mu\nu} \hat 1 \nonumber\\
     -\frac{(4-\omega)(\omega+1)}{18}
    R(-\Box)^{\omega-2}R \hat 1
    -\frac{2(2-\omega)(2\omega+1)}3\hat P(-\Box)^{\omega-2}R \nonumber\\
    +2(4\omega^2-1)\hat P(-\Box)^{\omega-2}\hat P +(2\omega+1)
    \hat{\cal R}_{\mu\nu}(-\Box)^{\omega-2}\hat{\cal R}^{\mu\nu}\Big\}+O[\Re^3],
\label{eq:qactgen}
\end{align}
where $\varGamma(z)$ is the Gamma function\footnote{Be aware of the difference in notation between the Gamma function $\varGamma(z)$ and the quantum action $\Gamma[\phi]$.}. The effective action \eqref{eq:qactgen} is the most general result valid for any gauge field, including gravitons, in arbitrary dimensions up to second order in curvature $\Re$. It is a functional of all background fields. Note that structures involving the Riemann curvature, such as $R_{\mu\nu\rho\sigma}f(\Box)R^{\mu\nu\rho\sigma}$, are not displayed in \eqref{eq:qactgen}. This happens because, due to the second Bianchi identity, the Riemann tensor satisfies a differential equation sourced by the Ricci tensor, thus the Riemann tensor can be determined in terms of the Ricci tensor up to boundary conditions. For asymptotically flat spaces, one can impose a trivial boundary condition such that the Riemann tensor vanishes for a Ricci-flat spacetime \cite{Barvinsky:1990up}. However, no such trivial condition can be imposed on asymptotically AdS or dS spaces. While one can still solve the differential equation for non-trivial boundary conditions (see Appendix~\ref{app}), for the time being we choose a more direct approach and leave the Riemann piece in the action below. This will also make the matching with the known results in the literature, such as the values in Table~\ref{table}, more transparent. In the next sections, however, we make use of the result in the Appendix~\ref{app} to ease our calculations.

Going back to Lorentzian signature and specializing to four dimensions $\omega=2$, the quantum action \eqref{eq:qactgen} for the mean metric $g_{\mu\nu} = \left<\hat g_{\mu\nu}\right>$ with vanishing mean matter fields becomes \cite{Donoghue:2014yha,Codello:2015mba}
\begin{equation}
\Gamma = \Gamma_\text{L}+\Gamma_\text{NL},
\label{action}
\end{equation}
where the local part reads
\begin{equation} \label{eq:localactionq}
        \Gamma_L = \int\mathrm{d}^4x \sqrt{-g} \left[ \frac{1}{16\pi G}\left(R-2\Lambda\right) + b_1 R^2
        + b_2 R_{\mu\nu} R^{\mu\nu} + b_3 R_{\mu\nu\rho\sigma}R^{\mu\nu\rho\sigma}\right],
\end{equation}
and the non-local one reads
\begin{align}
-\Gamma_{NL} &= \int\mathrm{d}^4x \sqrt{-g} \bigg[c_1 R \log\left(-\frac{\Box}{m^2}\right)R + c_2 R_{\mu\nu} 
\log\left(-\frac{\Box}{m^2}\right) R^{\mu\nu}\nonumber\\
& + c_3 R_{\mu\nu\rho\sigma} 
\log\left(-\frac{\Box}{m^2}\right) R^{\mu\nu\rho\sigma} \bigg],
\label{eq:nonlocalactionq}
\end{align}
where $m^2=-2\Lambda$ is the effective mass in the presence of a cosmological constant. We are assuming that $-\Box/m^2\gg 1$, in which case the form factor in the non-local piece of the action is dominated by $\log(-\Box/m^2)$. Other contributions, such as $-m^2/\Box$, are thus suppressed. The action \eqref{action} accounts for one-loop quantum corrections from both matter and gravitons running in the loops. A very useful way of dealing with non-local operators is via their spectral decomposition, which for the log is given by
\begin{equation}
\log\frac{-\Box}{m^2} = \int_0^\infty\mathrm{d}s\left(\frac{1}{m^2+s} - \frac{1}{-\Box+s}\right),
\label{eq:spec}
\end{equation}
where the second term above can be written in terms of the Green function of $-\Box + s$ and $s$ is just an integration variable with dimensions of mass (not to be confused with the proper time). Note that, differently from asymptotically flat spaces where non-local operators are determined by the direct substitution of retarded Green's function \cite{Barvinsky:1987uw}, the choice of the Green's function for non-local operators in spacetimes with boundary generally depends on the boundary conditions. In Euclidean signature, the classical solution is uniquely determined by its value at the conformal boundary and by the requirement of regularity at the horizon, which unambiguously lead to the retarded Green's function. Since transport coefficients must be computed in real time, the result is analitically continued to the Lorentzian signature at the end. On the other hand, the same aforementioned requirements are not sufficient to single out the Green's function by themselves if one works directly in Lorentzian signature, which in fact reflects the existence of various potential choices, such as the retarded, advanced and Feynman Green's function. The retarded Green's function in this case can be obtained by additionally imposing the incoming-wave boundary condition in which waves travel only towards the inside of the horizon. The choice of this additional boundary condition is also confirmed by the Schwinger-Keldysh formalism \cite{Herzog:2002pc}.

We stress that the non-local piece represents the infrared portion of quantum gravity, which is insensitive to the UV. This is in fact expected from the uncertainty principle: the higher the energy (or momentum) of the process, the smaller is the time (or space) window it occurs and vice versa. The coefficients $c_i$ are thus genuine predictions of the quantum theory of gravity. They are determined once the collection of fields $\Phi$ in \eqref{eq:pathint} and their respective spins are specified; see Table \ref{table} \footnote{The coefficients $c_i$ are universal predictions of quantum gravity irrespective of the background. In Ref. \cite{Donoghue:2014yha}, they were obtained by means of a perturbative series around Minkowski. Nonetheless, one arrives at the same values for $c_i$ using the background field method (with an arbitrary background) together with heat-kernel methods. In fact, the background is unspecified in Eq. (11) but its coefficients are fixed by the type and number of fields quantized in the path integral. The same values for $c_i$ were obtained in Ref. \cite{Codello:2015mba}, albeit presented in a different basis of curvature invariants, using this more general approach.}. The total contribution to each coefficient is given by simply summing the contribution from each field species.
\begin{table}[h!]
\centering 
\begin{tabular}{c c c c}
 \toprule
   & $c_1$ & $c_2$ & $c_3$ \\
  \midrule 
real scalar& $5(6 \xi -1)^2/(11520 \pi^2)$ &  $-2/(11520 \pi^2)$& $2/(11520 \pi^2) $ \\ 
 \midrule
Dirac spinor& $-5/(11520 \pi^2)$ & $ 8/(11520 \pi^2)$ & $7/(11520 \pi^2)$  \\ 
 \midrule
 vector & $ -50/(11520 \pi^2)$ & $176/(11520 \pi^2)$ & $ -26/(11520 \pi^2)$ \\ 
 \midrule
graviton & $430/(11520 \pi^2)$ & $-1444/(11520 \pi^2)$ & $424/(11520 \pi^2)$ \\ 
 \bottomrule 
\end{tabular}
\caption{Values of the coefficients $c_i$ for each spin ($\xi$ is the non-minimal coupling coefficient of scalars to gravity) extracted from \cite{Donoghue:2014yha}. Each value must be multiplied by the number of fields of its category present in the action $S[\Phi]$. The total value of each coefficient is then given by summing up all contributions.}
\label{table}
\end{table}
The local action, on the other hand, represents the high energy portion of quantum loops. As a result, the coefficients $b_i$ cannot be determined from first principles. They are renormalized parameters which must be fixed by observations or by matching with a UV completion. They satisfy the renormalization group equation
\begin{equation}
\mu\partial_\mu b_i = \beta_i
\label{eq:rg}
\end{equation}
for the beta functions $\beta_i = -2c_i$. Eq.~\eqref{eq:rg} is obtained after the renormalization of divergences from expressions of the form
\begin{equation}
b_i^B = b_i(\mu) - \frac{c_i}{\epsilon},\quad \frac{1}{\epsilon} = \frac{2}{4-d} - \gamma + \log(\sqrt{4\pi}).
\end{equation}
Note that the divergences are tied up to the finite contributions, ultimately leading to a relation between the $\beta_i$ and the coefficients of the non-local terms.

The effective action $\Gamma$ is a functional of the arbitrary mean field $g_{\mu\nu} = \left<\hat g_{\mu\nu}\right>$, which is not necessarily a solution of the classical equations of motion. In fact, $g_{\mu\nu}$ carries the information concerning the backreaction of the quantum fields integrated out in \eqref{eq:pathint}, thus describing the evolution of the background due to quantum fluctuations. When $g_{\mu\nu}$ does not satisfy the classical Einstein's equations, the quantum action is gauge independent \cite{DeWitt:1980jv}, but depends parametrically on the gauge fixing and on the parametrization of the quantum field. This issue has been solved by Vilkovisky by introducing a metric and a connection in the configuration space \cite{Vilkovisky:1984st}. Nonetheless, both the calculation of the shear viscosity and of the entropy involves the evaluation of the on-shell action in the AdS instanton
\begin{equation}
ds^2_\text{AdS} = -\left(\frac{r}{b}\right)^2 dt^2 + \frac{dr^2}{\left(\frac{r}{b}\right)^2} + \left(\frac{r}{b}\right)^2 d\vec x^2_2,
\label{eq:ads}
\end{equation}
where $b$ is the AdS radius, or in asymptotically AdS spaces, such as the SAdS black hole of mass $M$
\begin{equation}
ds^2_\text{SAdS} = - f(r) dt^2 + \frac{dr^2}{f(r)} + r^2 d\Omega_2^2,\quad f(r) = 1 - \frac{2M}{r} + \frac{r^2}{b^2},
\label{eq:sads}
\end{equation}
thus the parametrization and gauge fixing dependence will not be a concern for us \cite{DeWitt:1967ub,Grisaru:1975ei,Kallosh:1974yh}. We shall now see how to use the quantum action \eqref{action} to calculate the shear viscosity. In the rest of this paper we set $G=1$ for convenience.

\section{One-loop corrections to the shear viscosity}
\label{sec:eta}
The calculation of the shear viscosity of the gauge theory can be performed in many different ways \cite{Policastro:2001yc,Kovtun:2003wp,Policastro:2002se}. The most usual method employs the Kubo formula
\begin{equation}
\eta = -\lim_{\omega\to 0} \frac{1}{\omega} \operatorname{Im} G_R^{xy,xy}(\omega,\vec k=0),
\end{equation}
which relates the shear viscosity to the imaginary part of the retarded Green's function for the response of the $xy$ component of the energy-momentum tensor:
\begin{equation}
G_R^{xy,xy}(\omega,\vec k) = -i\int\mathrm{d}^3x\, e^{i\omega t - i\vec{k}\cdot \vec{x}}\theta(t)\left<\left[T^{xy}(t,\vec x),T^{xy}(0,\vec 0)\right]\right>,
\end{equation}
where $\theta(t)$ is the Heaviside step function. The Green function $G_R^{xy,xy}$ is then determined with the aid of the AdS/CFT correspondence $Z_\text{gauge} = Z_\text{AdS}$, which translates into the GKP--Witten relation in real time
\begin{equation}
\left<\exp\left({i\int\mathrm{d}^3x\, h_{xy}^{(0)} T^{xy}}\right)\right> = \exp\left({i\Gamma[h_{xy}^{(0)}]}\right),
\label{eq:gkp}
\end{equation}
where $h_{xy}^{(0)} = h_{xy}|_{r=\infty}$ denotes the gravitational perturbation polarized parallel to the brane at the AdS boundary and $\Gamma[h_{xy}^{(0)}]$ is the on-shell quantum action \eqref{action} for the gravitational perturbation in AdS. Under functional variations, one can find from \eqref{eq:gkp} all the correlation functions of the gauge theory. In particular, the one-point function $\left<T^{xy}\right> = -G_R^{xy,xy}h_{xy}^{(0)}$ is given by
\begin{equation}
\left<T^{xy}\right> = \frac{\delta\Gamma[h_{xy}^{(0)}]}{\delta h_{xy}^{(0)}}.
\end{equation}
Finding the the shear viscosity $\eta$ thus amounts on the calculation of the dynamics of the component $h_{xy}$ of the bulk perturbation. We shall now see how the one-loop correction in $\Gamma$ affects its evolution.

In the realm of effective field theory, the quantum action has to be treated perturbatively. The lowest order determines the degrees of freedom and their interactions, while higher order terms make contributions only to the latter, i.e. to vertices of Feynman diagrams. This is the standard lore of effective field theory. We thus need to linearize the quantum action around some fixed background by performing the transformation $g_{\mu\nu} = \bar g_{\mu\nu} + h_{\mu\nu}$, where $\bar g_{\mu\nu}$ is the background metric. The quantum action then becomes
\begin{equation}
\Gamma[\bar g + h] = \Gamma[\bar g] + \int\mathrm{d}^4x\frac{\delta \Gamma^{(1)}[\bar g]}{\delta g^{\mu\nu}(x)}h^{\mu\nu}(x) +\frac12 \iint\mathrm{d}^4x\mathrm{d}^4y \frac{\delta^2 \Gamma^{(0)}[\bar g]}{\delta g^{\mu\nu}(x)g^{\rho\sigma}(y)} h^{\mu\nu}(x)h^{\rho\sigma}(y) + \cdots,
\label{eq:qtactexp}
\end{equation}
where we made a loop expansion $\Gamma = \sum_n \Gamma^{(n)}$, denoting $\Gamma^{(0)}$ as the classical Einstein-Hilbert action with a cosmological constant and $\Gamma^{(1)}$ as the one-loop contribution. Note that the $\Gamma^{(0)}$ does not contribute to the linear term because the linearization is around the AdS instanton, thus the variation of $\Gamma^{(0)}$ vanishes by the classical equations of motion. The radiative correction $\Gamma^{(1)}$ does not contribute to the quadratic order either as it is suppressed in this perturbative treatment.

To calculate $\delta^2\Gamma^{(0)}[\bar g]$ we need to compute the variation of the logarithm operator. From Eq.~\eqref{eq:spec}, we have
\beq
\delta\log\left(-\frac{\Box}{m^2}\right) = \int_0^\infty\mathrm{d}s\, \delta\frac{1}{\Box - s}.
\label{eq:varlog}
\eeq
The variation in the integrand is formally given by \cite{DeWitt:1965jb}
\beq
\delta \frac{1}{\Box - s} = -G(x,x';s) \delta(\Box) G(x,x';s),
\label{eq:varG}
\eeq
where $G(x,x';s)$ is the Green's function of $\Box - s$. Adopting Riemann normal coordinates, we can write the Green's function in curved spaces as a curvature expansion \cite{Bunch:1979uk}
\beq
G(x,x';s) = G^\eta(x,x';s) + \mathcal{O}(\bar R),
\label{eq:normalcoord}
\eeq
where $G^\eta$ is the Green's function in flat spacetime. Because of $\delta\Box$, additional powers of curvature from \eqref{eq:normalcoord} only give higher-order contributions to \eqref{eq:varG}. From Eqs.~\eqref{eq:varlog}, \eqref{eq:varG} and \eqref{eq:normalcoord}, we then find
\beq
\delta\log\left(-\frac{\Box}{m^2}\right) = \int_0^\infty\mathrm{d}s\, G^\eta(x,x';s) \delta(\Box) G^\eta(x,x';s) = \mathcal{O}(\bar R).
\eeq
Therefore, terms such as
\beq
\bar R\,\delta\left[\log\left(-\frac{\Box}{m^2}\right)\right]\bar R = \mathcal{O}(\bar R^3),
\label{eq:termlog}
\eeq
can be neglected to second order in curvature \cite{Donoghue:2014yha,Donoghue:2015nba}. Using \eqref{eq:termlog}, we can then obtain $\delta^2\Gamma^{(0)}[\bar g]$ in \eqref{eq:qtactexp}, which ultimately leads to the equations of motion for $h_{\mu\nu}$ \cite{Kuntz:2017pjd}:
\beq
\frac{1}{16\pi G}\Box\bar h_{\mu\nu} = \Delta G^{L}_{\mu\nu} + \Delta G^{NL}_{\mu\nu},
\label{eq:perteom}
\eeq
where the local part reads
\begin{align}
\Delta G^{L}_{\mu\nu} &= (b_1 - b_3)\left[2\left(\bar R_{\mu\nu}-\frac{1}{4}\bar g_{\mu\nu}\bar R\right)\!\bar R-2 \left(\nabla_{\mu}\nabla_{\nu}-\bar g_{\mu\nu}\square\right) \bar R\right]\nonumber\\
& -(b_2+4b_3) \bigg[-\frac12 g_{\mu\nu}\bar R_{\rho\sigma}\bar R^{\rho\sigma} + \Box\bar R_{\mu\nu} + g_{\mu\nu}\nabla_\rho\nabla_\sigma\bar R^{\rho\sigma}\nonumber\\
& + \bar R_\mu^\sigma\bar R_{\nu\sigma} + \bar R_\nu^\sigma\bar R_{\mu\sigma} - \nabla_\rho\nabla_\mu \bar R^{\rho}_{\nu} - \nabla_\rho\nabla_\nu \bar R_{\mu}^{\rho}\bigg],
\end{align}
and the non-local one reads
\begin{align}
&\Delta G^{NL}_{\mu\nu} = (c_1 - c_3)\left[2\left(\bar R_{\mu\nu}-\frac{1}{4}\bar g_{\mu\nu}\bar R\right)\!\log\left(-\frac{\square}{\mu^{2}}\right)\bar R-2 \left(\nabla_{\mu}\nabla_{\nu}-\bar g_{\mu\nu}\square\right) \log\left(-\frac{\square}{\mu^{2}}\right)\bar R\right]\nonumber\\
& -(c_2+4c_3) \bigg[-\frac12 g_{\mu\nu}\bar R_{\rho\sigma}\log\left(-\frac{\Box}{\mu^2}\right)\bar R^{\rho\sigma} + \Box\log\left(-\frac{\Box}{\mu^2}\right)\bar R_{\mu\nu} + g_{\mu\nu}\nabla_\rho\nabla_\sigma\log\left(-\frac{\Box}{\mu^2}\right)\bar R^{\rho\sigma}\nonumber\\
& + \bar R_\mu^\sigma\log\left(-\frac{\Box}{\mu^2}\right)\bar R_{\nu\sigma} + \bar R_\nu^\sigma\log\left(-\frac{\Box}{\mu^2}\right)\bar R_{\mu\sigma} - \nabla_\rho\nabla_\mu \log\left(-\frac{\Box}{\mu^2}\right)\bar R^{\rho}_{\nu} - \nabla_\rho\nabla_\nu \log\left(-\frac{\Box}{\mu^2}\right)\bar R_{\mu}^{\rho}\bigg],
\end{align}
with $\bar h_{\mu\nu} = h_{\mu\nu} -\frac12 \bar g_{\mu\nu} h$. All covariant derivatives above are constructed with $\bar g_{\mu\nu}$. Note that, for both AdS \eqref{eq:ads} and SAdS \eqref{eq:sads}, the metric $\bar g_{\mu\nu}$ and the Ricci tensor $\bar R_{\mu\nu} = \Lambda \bar g_{\mu\nu}$ are diagonal. Using this fact, together with $\nabla_\rho \bar g_{\mu\nu} = 0$ and $\partial_\mu \Lambda = 0$, we can reduce the equation of motion \eqref{eq:perteom} for $h_{xy}$ to\footnote{Note that even the terms with cross contractions, such as $R_\mu^\sigma\log\left(-\Box/\mu^2\right)R_{\nu\sigma} = \log\left(M_c^2/\mu^2\right)\Lambda^2 g_{\mu\nu}$, turn out to be diagonal after the regularization used in Eq.~\eqref{eq:logreg} below.}
\begin{equation}
\Box \bar h_{xy} = 0.
\end{equation}
We conclude that, at least at one-loop order, the evolution of $h_{xy}$ is exactly as in classical general relativity. The on-shell action also takes the same form as in general relativity
\beq
\Gamma = \int\mathrm{d}^3x \left.\frac{1}{2u^3}\phi \phi'\right|_{u=0}, \quad \text{with}\quad \phi = g^{xx}h_{xy},
\eeq
since in the effective field theory approach the higher-derivative terms do not contribute to the kinematical (i.e. bilinear) part of the action of $h_{\mu\nu}$ (see Eq.~\eqref{eq:qtactexp}), thus they do not lead to new boundary terms upon integration by parts. Boundary counterterms in the holographic renormalization of $h_{xy}$ are consequentely not affected by the higher-derivative terms and holographic renormalization follows the same procedure as for a free scalar field \cite{deHaro:2000vlm}.

The shear viscosity is then given by the classical result \cite{Klebanov:1997kc,Gubser:1997yh,Kovtun:2004de,Das:1996we,Emparan:1997iv}
\begin{equation}
\label{eq:eta}
\eta = \frac{\sigma_\text{abs}(\omega=0)}{16\pi},
\end{equation}
where the absorption cross-section $\sigma_\text{abs}(\omega=0) = A_+$ equals the horizon area.

\section{One-loop corrections to the entropy}
\label{sec:s}
In this section, we use the Euclidean method to calculate the entropy of the SAdS black hole, following the usual procedure of transforming to the Euclidean time $\tau = it$, for $\tau\in (0,\beta)$, and imposing periodic boundary conditions so to avoid conical singularities. In asymptotically AdS spaces, one can use the canonical ensemble to compute thermodynamical quantities, where the black hole is put in contact with a thermal bath. To get rid of potential divergences in the entropy, one must subtract the AdS entropy from the SAdS one, which corresponds to normalizing the partition function with respect to AdS\footnote{Alternatively, the AdS on-shell action can be made automatically finite by including all the necessary boundary counterterms for holographic renormalization \cite{Papadimitriou:2005ii}. Although holographic renormalization is much more powerful and general, background subtraction is enough for our purposes.}. This has the effect of eliminating the contribution from the thermal bath.

In Euclidean time, the normalized partition function reads
\begin{equation}
Z(\beta) = e^{-\Delta\Gamma},
\end{equation}
where $\Delta\Gamma\equiv \Gamma_\text{SAdS}-\Gamma_\text{AdS}$ is the difference of the on-shell actions evaluated at SAdS and AdS, respectively. Differently from the standard Schwarzschild case, where the contribution to the partition function comes solely from the Gibbons-Hawking-York boundary terms, we can disregard any boundary terms\footnote{The effective action in spaces with boundary was studied in \cite{Barvinsky:1995dp}, where a boundary term analogous to the Gibbons-Hawking one was found. This boundary term modifies the one-loop quantum corrections to give a correct balance between the volume and boundary parts for the effective action. This means that the boundary UV divergences do not require independent renormalization and are automatically renormalized simultaneously with their volume part.} because they are canceled out in the difference $\Gamma_\text{SAdS}-\Gamma_\text{AdS}$.

Let us now calculate the on-shell action. Note that both AdS and SAdS metrics satisfy the classical equations
\begin{align}
\label{eq:eomcl1}
R_{\mu\nu} &= \Lambda g_{\mu\nu},\\
\label{eq:eomcl2}
R &= 4\Lambda.
\end{align}
The main difficulty is then to calculate the action of the log operator on the metric and on the cosmological constant. Using the spectral representation of the log, we find
\begin{align}
\log\left(-\frac{\Box}{m^2}\right)g_{\mu\nu} &= \int_{M_c^2}^\infty\mathrm{d}s\left[\frac{1}{s+m^2} - \frac{1}{-\Box+s}\right]g_{\mu\nu}\nonumber\\
&= \int_{M_c^2}^\infty\mathrm{d}s\left[\frac{1}{s+m^2} - \frac{1}{s}\right]g_{\mu\nu}\nonumber\\
&= \log\left(\frac{M_c^2}{m^2}\right)g_{\mu\nu},
\label{eq:logreg}
\end{align}
where $M_c$ is some mass scale of $g_{\mu\nu}$ used to regulate the divergence appearing at $s=0$. For SAdS, we take $M_c = M$ as the mass of the black hole, while for AdS we take $M_c=\varepsilon\to 0$ as a temporary cut-off. The final result will turn out to be independent of $\varepsilon$. The action on the cosmological constant gives the same result as in Eq. \eqref{eq:logreg} for obvious reasons. The Box operator disappears in the second equality above because of the metric compatibility of the connection, which implies $\Box g_{\mu\nu} = 0$. Thus,
\begin{equation}
(-\Box + s) g_{\mu\nu} = s g_{\mu\nu} \iff \frac{1}{-\Box + s} g_{\mu\nu} = \frac{1}{s} g_{\mu\nu} + \frac{1}{s}\text{BC},
\label{bc}
\end{equation}
where BC denotes the boundary condition imposed on the non-local operator $1/(-\Box + s)$. In obtaining \eqref{eq:logreg}, we have formally disregarded BC by choosing the trivial boundary condition $\text{BC} = 0$.

We stress that the results do not depend on $\varepsilon$, thus one can remove such a cutoff at the end without affecting our results. On the other hand, the results do depend on the SAdS mass $M_c = M$. This is nonetheless expected as we now explain. Due to the spectral representation of non-local operators, one can make analogies with the K\"allen-Lehmann representation of full propagators even though $\log(- \Box / m^2)$ is not the propagator of the theory in question. In the K\"allen-Lehmann representation, one-particle states are isolated and the integral over the spectral density starts from the mass of the lightest multi-particle state. In the spectrum of $\log(- \Box / m^2)$, no ``one-particle state'' (or isolated poles) exists below $M$ and above $M$ the multi-particle spectra (microstates) of the black hole is formed in accordance to the holographic principle. It is thus natural to expect the integral in \eqref{eq:logreg} to start from $s=M$. 
Mathematically, the need of such a cutoff reflects the formal choice of boundary condition of the non-local operator ($\text{BC} = 0$). In fact, cutting off the lower limit of the integral by $M_c$ and setting $\text{BC}=0$ is equivalent to choosing
\beq
\text{BC} = -\frac{M_c^2}{M_c^2 + s^2} g_{\mu\nu}
\eeq
in Eq. \eqref{bc} and letting the lower limit of the integral \eqref{eq:logreg} start from zero. Note that one is not removing the non-local contribution because $\log(M_c^2 / m^2)$ is the actual result of such a contribution. Moreover, $\log(- \Box / m^2) = \log(M_c^2 / m^2)$ only when the non-local operator is applied to the metric. In this case, one can think of $\log(M_c^2 / m^2)$ as an infinite-dimensional eigenvalue of the non-local operator $\log(- \Box / m^2)$ and $g_{\mu\nu}$ as its corresponding eigenvector.

Using Eqs. \eqref{eq:eomcl1}, \eqref{eq:eomcl2} and \eqref{eq:logreg} in \eqref{eq:nonlocalactionq}, gives
\begin{align}
\Gamma_\text{SAdS} &= \left[-\Lambda + \frac{4\Lambda^2}{3}\log\left(\frac{M^2}{m^2}\right)(12c_1+3c_2+2c_3)\right]V_\text{SAdS}\\
\Gamma_\text{AdS} &= \left[-\Lambda + \frac{4\Lambda^2}{3}\log\left(\frac{M_c^2}{m_1^2}\right)(12c_1+3c_2+2c_3)\right]V_\text{AdS},
\end{align}
where $m_1^2 = -2\Lambda_1$ is the arbitrary effective mass of AdS and
\begin{align}
V_\text{SAdS} &= \frac{4\pi}{3}\beta (L^3-r_+^3),\\
V_\text{AdS} &= \frac{4\pi}{3}\beta_1 L^3,
\end{align}
are the volume of SAdS and AdS, respectively, where we have introduced an infrared cut-off $L$ and $r_+$ is the horizon radius obtained by solving $f(r_+)=0$. The period $\beta$ of SAdS is fixed so to avoid the conical singularity
\begin{equation}
\beta = \frac{4\pi b^2 r_+}{b^2 + 3r_+^2}.
\end{equation}
On the other hand, the period $\beta_1$ of AdS is \textit{a priori} arbitrary. However, since the two metrics must coincide at $L\to\infty$, the time coordinate must have the same period:
\begin{equation}
\beta_1 \sqrt{1+\frac{L^2}{b^2}} = \beta\sqrt{1-\frac{2M}{L}+\frac{L^2}{b^2}} \implies \frac{\beta_1}{\beta} \approx 1 - \frac{Mb^2}{L^3}.
\end{equation}
Therefore, the difference of the on-shell actions $\Delta \Gamma\equiv \Gamma_\text{SAdS}-\Gamma_\text{AdS}$ reads
\begin{equation}
\Delta \Gamma = \Delta\Gamma^{(0)} + \Delta\Gamma^{(1)},
\end{equation}
where
\begin{equation}
\Delta\Gamma^{(0)} = \frac{\pi r_+^2 (b^2 - r_+^2)}{b^2 + 3r_+^2}
\end{equation}
is the usual general relativistic result and
\begin{equation}
\label{eq:dact1}
\Delta\Gamma^{(1)} = \frac{16\pi}{9}\Lambda^2(12c_1 + 3c_2 + 2c_3)\beta \left[\log\left(\frac{M^2}{m^2}\right)(L^3-r_+^3) - \log\left(\frac{M_c^2}{m_1^2}\right)(L^3-Mb^2)\right]
\end{equation}
is the one-loop contribution. The divergence $L\to\infty$ in $\Delta\Gamma^{(1)}$ is not automatically removed as in the classical part $\Delta\Gamma^{(0)}$, but we can exploit the arbitrariness of the effective mass $m_1$ to cancel out this divergence. This is achieved with the choice $m_1^2 = \frac{M_c^2}{M^2}m^2$, which makes the logarithms in Eq. \eqref{eq:dact1} equal and ultimately leads to
\begin{equation}
\Delta\Gamma^{(1)} = \frac{96\pi^2}{3}(12c_1+3c_2+2c_3)\left[\log\left(\frac{r_+^2}{m^2}\right)+2\log\left(1+\frac{r_+^2}{b^2}\right)\right]\frac{r_+^2(b^2-r_+^2)}{b^2(b^2+3r_+^2)}.
\end{equation}
The entropy is finally given by
\begin{align}
S &= (\beta\partial_\beta - 1)\Delta\Gamma\nonumber\\
&= \frac{A_+}{4} + \frac{8\pi(12c_1+3c_2+2c_3)A_+}{b^2}\log\left(\frac{M^2}{m^2}\right) + \Xi\left(\frac{r_+}{b}\right),
\label{eq:s}
\end{align}
where $A_+ = 4\pi r_+^2$ is the horizon area and
\begin{equation}
\Xi\left(\frac{r_+}{b}\right) = 64\pi^2 (12c_1+3c_2+2c_3) \frac{r_+^2}{b^2}\frac{(1-r_+^2/b^2)(1+3r_+^2/b^2)}{(1+r_+^2/b^2)(1-3r_+^2/b^2)}.
\end{equation}
The logarithmic correction to the entropy seems to be a universal feature of quantum gravity \cite{Aros:2010jb,Carlip:2000nv}. It has been obtained in different contexts using different techniques \cite{Banerjee:2008fz,Cai:2009ua,Aros:2010jb,Banerjee:2010qc,Banerjee:2011jp,Sen:2012cj,Sen:2012dw,Kaul:2000kf,Carlip:2000nv,Govindarajan:2001ee,Mukherji:2002de,Solodukhin:1994st,Fursaev:1994te,Mann:1997hm}. Our result is yet another instance of this apparent universality.

We are finally able to calculate the shear viscosity to entropy ratio, which is the main concern of this paper. Combining Eqs. \eqref{eq:eta} and \eqref{eq:s}, we obtain
\begin{equation}
\frac{\eta}{s} = \frac{1}{4\pi}\left[1+\frac{32\pi(12c_1+3c_2+2c_3)}{b^2}\log\left(\frac{M^2}{m^2}\right) + \frac{4}{A_+}\Xi\left(\frac{r_+}{b}\right)\right]^{-1}.
\label{eq:r1}
\end{equation}
Eq. \eqref{eq:r1} has been obtained for SAdS with spherical horizon. Note that when the AdS black hole has a spherical horizon, the dual gauge theory lives on a sphere. Hydrodynamics on a sphere is much more complicated because the existence of gapless excitations depends on the hierarchy of scales, such as the mean-free-path, that now also include the radius of the space. For the simpler case of a black brane with planar horizon, which can be seen as the limiting case of a large black hole with constant $r_+$, the above result simplifies to
\begin{equation}
\frac{\eta}{s} = \frac{1}{4\pi}\left[1+\frac{32\pi(12c_1+3c_2+2c_3)}{b^2}\log\left(\frac{M^2}{m^2}\right)\right]^{-1}.
\end{equation}
For completeness, we include the result for the local action even though it does not constitute a prediction of the IR theory:
\begin{equation}
\frac{\eta}{s} = \frac{1}{4\pi}\left\{1+\frac{32\pi}{b^2}\left[(12c_1+3c_2+2c_3)\log\left(\frac{M^2}{m^2}\right)+(12b_1+3b_2+2b_3)\right]\right\}^{-1}.
\label{eq:etasfull}
\end{equation}
To put our result in a form that can be interpreted from the gauge side of the correspondence, we first recall that the coefficients $b_i = b_i(\mu)$ run with the renormalization scale $\mu$. The unphysical scale $\mu$ must not show up in the observables. Requiring that the 1PI action is $\mu$-independent forces $\mu=m$, thus the coefficients $b_i$ turn out to run with the effective mass $m^2=-2\Lambda$. One can then use dimensional transmutation to trade the dimensionless coupling constants $b_i$ with a dimensionful energy scale $E_{UV}$:
\begin{equation}
12b_1+3b_2+2b_3 = -(12c_1+3c_2+2c_3)\log\left(\frac{M^2}{E_{UV}^2}\right).
\label{eq:trans}
\end{equation}
The fact that $b_i$ are unknown UV parameters are now translated into an unknown UV scale, which must be fixed by observations or by matching with a UV completed theory. In addition, the AdS/CFT dictionary in four dimensions implies
\begin{equation}
N^2 \sim \frac{b^2}{G},
\label{eq:dict}
\end{equation}
where $b$ is the AdS radius and $G$ is the four-dimensional Newton's constant. If we reinstate $G$ (it was initially set to unity) in Eq.~\eqref{eq:etasfull}, we can see that $\eta/s$ only depends on $G$ and $b$ through the combination $G/b^2$. Therefore, from Eqs. \eqref{eq:trans} and \eqref{eq:dict}, we find
\begin{equation}
\frac{\eta}{s} = \frac{1}{4\pi}\left[ 1 + \frac{32\pi}{N^2} (12c_1+3c_2+2c_3) \log\left(\frac{M^2}{E_{UV}^2}\right) \right]^{-1}.
\label{finaletas}
\end{equation}
We conclude that the quantum gravitational corrections indeed correspond to finite $N$ corrections on the gauge theory side, while $M$ is the total energy of the gauge physical system and $E_{UV}$ is a UV mass scale to be fixed by observations. From Eq. \eqref{eq:dict}, one can see that $N\to\infty$ corresponds to a vanishing Newton's constant, making the Einstein-Hilbert term (i.e. classical gravity) arbitrarily large, thus suppressing the local and non-local higher curvature terms. Thus, the fact that the corrections in Eq. \eqref{finaletas} vanish for $N\to\infty$ is a good check of self-consistency as the classical result must indeed be recovered in this limit.

Note that Eq. \eqref{finaletas} is unlikely to be a good fit for the common substances (e.g. water, helium, nitrogen) for all temperatures as the effective field theory breaks down for large (Planckian) energies (or equivalently large temperatures). The precision can nonetheless be improved by incorporating higher loop corrections. That being said, for any given temperature $T$ one can solve Eq. \eqref{finaletas} for the coefficient $\alpha \equiv \frac{32\pi}{N^2} (12 c_1 + 3 c_2 + 2 c_3)$, leading to:
\[
\alpha = \frac{1 - 4\pi (\eta/s)}{4\pi\,(\eta/s) \log(M^2(T)/E_{UV}^2)},
\]
where $M(T)$ is obtained by solving the system
\begin{align*}
1 - \frac{2 M}{r_+} + \frac{r_+^2}{b^2} = 0,\\
T = \frac{b^2 + 3 r_+^2}{4\pi b^2 r_+}.
\end{align*}
Thus, for any data point $(T, \eta/s)$ of any substance, one can in principle find the combination of coefficients $\alpha$ to fit the result. Note that the coefficientes $c_i$ are accompanied by the number of particles in the theory, thus one has a lot of freedom to choose $\alpha$. Once again, we stress that despite the ability of fitting an arbitrary data point, the effective field theory is only reliable for small temperatures.

The crucial aspect of Eq. \eqref{finaletas} is that the combination of coefficients $12c_1+3c_2+2c_3$ can be either positive or negative depending on the spin of particles which had been integrated out to obtain the quantum action (see Table \ref{table}). Therefore, the KSS bound
\begin{equation}
\frac{\eta}{s} > \frac{1}{4\pi}
\end{equation}
is not necessarily satisfied by all kinds of integrated particles and does not seem to represent a fundamental bound that holds beyond the classical level. We must stress that, contrary to other violations of the KSS bound \cite{Buchel:2004di,Buchel:2008ac,Buchel:2008wy,Myers:2008yi,Buchel:2008ae,Garousi:2008ai,Ghodsi:2009hg,Buchel:2008sh,Cohen:2007qr,Son:2007xw,Brigante:2007nu,Brigante:2008gz,Kats:2007mq}, the above result has been obtained within general relativity by using effective field theory techniques to identify the infrared portion of quantum gravity, which permitted the evaluation of the one-loop contribution to $\eta/s$. The quantum action \eqref{action} is not supposed to be seen as a modification of gravity, after all the degrees of freedom and the interactions are the ones of general relativity, but the latter receives radiative corrections due to quantum fields running in the loops.

\section{Conclusions}
\label{sec:conc}
In this paper, we have studied one-loop corrections of matter fields and gravitons to the shear viscosity to entropy ratio in AdS$_4$/CFT$_3$. Although the former does not receive any correction at one-loop order, the latter gets corrected by a term proportional to the logarithm of the black hole mass. The coefficient of this correction does not have a definite sign because of its spin dependence. We thus argued that the celebrated KSS bound cannot be seen as a fundamental relation beyond the tree level. We should emphasize, once again, that our result has been obtained within general relativity by using effective field theory to calculate the leading order corrections. The aforementioned violation is entirely due to the quantum nature of fields, including the graviton excitations, in the SAdS background. 

\noindent{\it Acknowledgments:}
IK has the work supported by the National Council for Scientific and Technological Development -- CNPq (Brazil) under grant number 155342/2018-5. RdR is grateful to FAPESP (Grant No. 2017/18897-8), to CNPq (Grants No. 303390/2019-0, No. 406134/2018-9 and No. 303293/2015-2) and to HECAP - ICTP, Trieste, for partial financial support, and this last one also for the hospitality.

\appendix
\section{Elimination of the Riemann tensor for spaces with $\Lambda\neq0$}
\label{app}
It was shown in \cite{Barvinsky:1990up} that the Riemann tensor can be eliminated from the quantum action for asympotically flat spaces. We generalize the argument to asympotically AdS and dS spaces. The second Bianchi identity reads
\begin{equation}
\nabla^\lambda R^{\alpha\beta\mu\nu} + \nabla^\nu R^{\alpha\beta\lambda\mu} + \nabla^\mu R^{\alpha\beta\nu\lambda} = 0,
\label{eq:b1}
\end{equation}
which can be contracted to give
\begin{equation}
\nabla_\alpha R^{\alpha\beta\mu\nu} = \nabla^\mu R^{\beta\nu} - \nabla^\nu R^{\beta\mu}.
\label{eq:b2}
\end{equation}
Contracting \eqref{eq:b1} with $\nabla_\lambda$ and using \eqref{eq:b2} with the aid of the commutation of covariant derivatives gives
\begin{align}
\Box R^{\alpha\beta\mu\nu} &= \nabla^\mu\nabla^\alpha R^{\nu\beta} - \nabla^\nu\nabla^\alpha R^{\mu\beta} - \nabla^\mu\nabla^\beta R^{\nu\alpha} + \nabla^\nu\nabla^\beta R^{\mu\alpha}\nonumber\\
& -4R^{\alpha\ [\mu}_{\ \sigma\ \lambda}R^{\beta\sigma\nu]\lambda} + 2R^{[\mu}_\lambda R^{\alpha\beta\lambda\nu]} - R^{\alpha\beta}_{\ \ \sigma\lambda}R^{\mu\nu\sigma\lambda}.
\label{eq:b3}
\end{align}
Eq. \eqref{eq:b3} can be solved iteratively for the Riemann tensor, which is determined in terms of the Ricci tensor up to boundary conditions. For asymptotically AdS or dS spaces, one can use a maximally symmetric space as the boundary condition such that
\begin{equation}
R_{\mu\nu\rho\sigma} = \frac{\Lambda}{3}(g_{\mu\rho} g_{\nu\sigma} - g_{\mu\sigma} g_{\nu\rho})\quad\text{for}\quad R_{\mu\nu} = \Lambda g_{\mu\nu}.
\label{eq:bc}
\end{equation}
Therefore, the Riemann tensor can be uniquely determined from Eq. \eqref{eq:b3} by imposing the boundary condition \eqref{eq:bc} and using some appropriate Green function for $\Box^{-1}$. To lowest order, one finds
\begin{align}
R^{\alpha\beta\mu\nu} &= \nabla^\mu\nabla^\alpha\Box^{-1} R^{\nu\beta} - \nabla^\nu\nabla^\alpha\Box^{-1} R^{\mu\beta} - \nabla^\mu\nabla^\beta\Box^{-1} R^{\nu\alpha} + \nabla^\nu\nabla^\beta\Box^{-1} R^{\mu\alpha}\nonumber\\
& + \frac{\Lambda}{3}(g^{\alpha\mu} g^{\beta\nu} - g^{\alpha\nu} g^{\beta\mu}) + \mathcal{O}(R^2),
\end{align}
which can be used to eliminate the Riemann tensor from the effective action in favor of the Ricci tensor and the cosmological constant. In particular, the computation of the square of the Riemann tensor gives
\begin{equation}
R_{\mu\nu\alpha\beta}R^{\mu\nu\alpha\beta} = 4R_{\mu\nu}R^{\mu\nu} - R^2 + \frac{8\Lambda^2}{3} + \nabla_\mu\zeta^\mu + \mathcal O(R^3),
\label{eq:r2}
\end{equation}
where
\begin{align}
\zeta^\mu &= 4\nabla^\nu\Box^{-1}R^{\alpha\beta}\left(\nabla_\nu\nabla^\mu\Box^{-1}R_{\alpha\beta} - \nabla_\beta\nabla^\mu\Box^{-1} R_{\nu\alpha} + \nabla_\alpha\nabla_\beta\Box^{-1}R^\mu_\nu - \nabla_\nu\nabla_\alpha\Box^{-1}R^\mu_\beta\right)\nonumber\\
& +4\Box^{-1}R_{\alpha\beta}\nabla^\alpha R^{\beta\mu} - 4R^{\alpha\beta}\nabla^\mu\Box^{-1}R_{\alpha\beta} + R\nabla^\mu\Box^{-1}R - 2\Box^{-1}R^{\mu\nu}\nabla_\nu R.
\end{align}
Note that one recovers the result of \cite{Barvinsky:1990up} for $\Lambda = 0$. While the third term on the RHS of \eqref{eq:r2} does not contribute to the equations of motion \eqref{eq:perteom}, it definitely contributes to the entropy \eqref{eq:s}.
 

\bigskip{}

\end{document}